\documentclass[nohyper]{PoS}
\pdfoutput=1

\usepackage{cite}
\usepackage{amsmath}
\usepackage{amssymb}
\usepackage{revsymb}
\usepackage{url}

\newcommand{\beq}{\begin{equation}}
\newcommand{\eeq}{\end{equation}}
\newcommand{\bea}{\begin{eqnarray}}
\newcommand{\eea}{\end{eqnarray}}

\title{Flux tubes in QCD with (2+1) HISQ fermions}

\ShortTitle{Flux tubes in QCD with (2+1) HISQ fermions}

\author{\speaker{Leonardo Cosmai}%
\\
        INFN - Sezione di Bari, I-70126 Bari, Italy\\
        E-mail: \email{leonardo.cosmai@ba.infn.it}}
\author{Paolo Cea\\
            Dipartimento di Fisica dell'Universit\`a di Bari, I-70126 Bari, Italy and INFN, Sezione di Bari, I-70126 Bari, Italy\\
            E-mail: \email{paolo.cea@ba.infn.it}}
\author{Francesca Cuteri\\
            Institut f\"ur Theoretische Physik, Goethe Universit\"at,
            60438 Frankfurt am Main, Germany\\
            E-mail: \email{cuteri@th.physik.uni-frankfurt.de}}
\author{Alessandro  Papa\\
            Dipartimento di Fisica, Universit\`a della Calabria, \\
            \& INFN - Gruppo Collegato di Cosenza, I-87036 Rende, Italy \\
            E-mail: \email{papa@cs.infn.it}}

\abstract{We investigate the transverse profile of the chromoelectric field generated by a quark-antiquark pair in the vacuum of (2+1) flavor QCD. 
Monte Carlo simulations are performed adopting the HISQ/tree action discretization, as implemented in the publicly available MILC code, 
suitably modified to measure the chromoelectric field. 
We work on the line of constant physics, with physical strange quark mass $m_s$ and light to strange mass ratio $m_l/m_s = 1/20$.}

\FullConference{34th annual International Symposium on Lattice Field Theory\\
		24-30 July 2016\\
		University of Southampton, UK}

\begin{document}

\section{Introduction}
\label{introd}

\begin{figure}[b] 
\centering
\includegraphics[scale=0.4,clip]{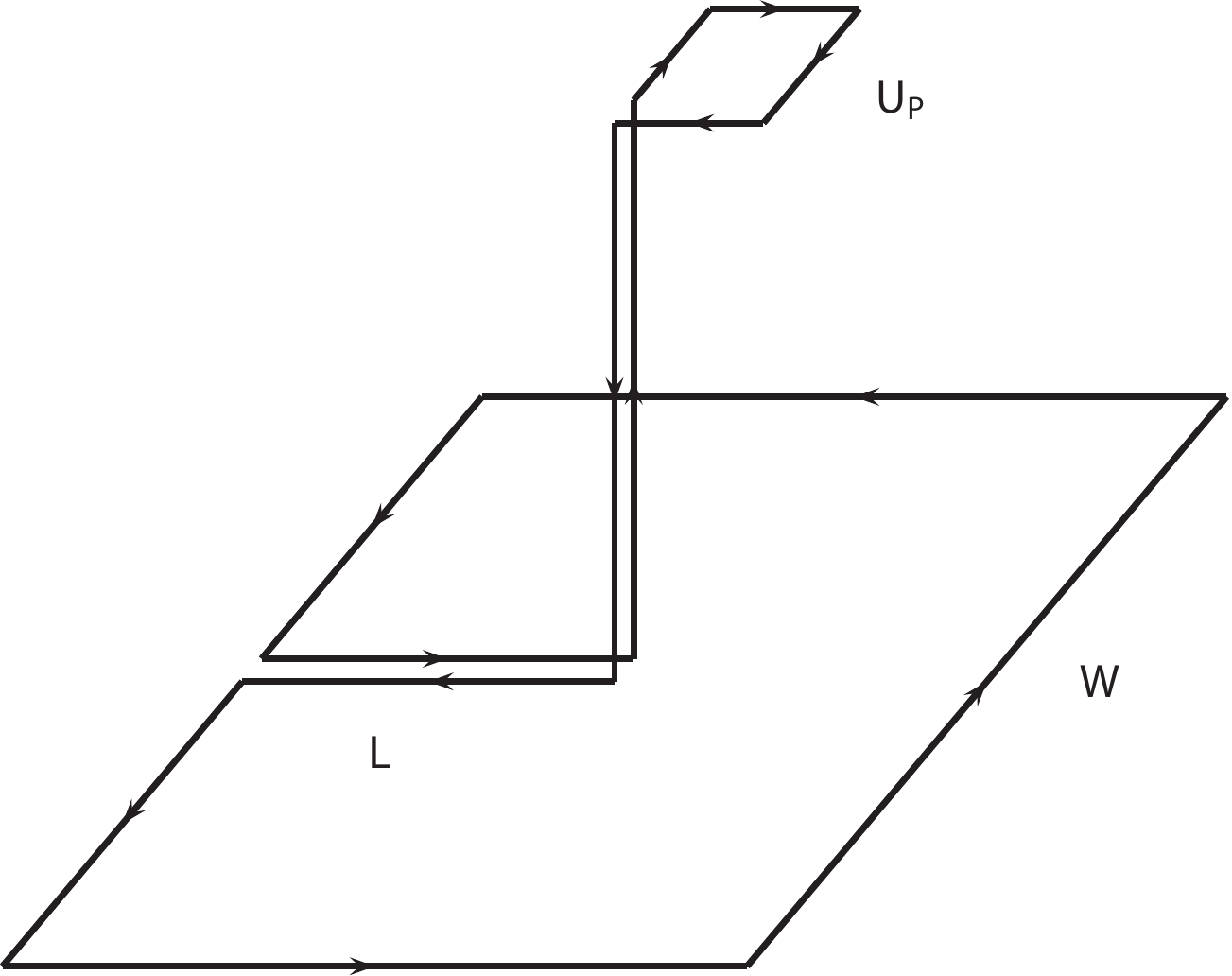} 
\caption{The connected correlator given in Eq.~(\protect\ref{rhoW})
between the plaquette $U_{P}$ and the Wilson loop
(subtraction in $\rho_{W}^{\rm conn}$ not explicitly drawn).}
\label{fig:op_W}
\end{figure}

Many fundamental questions are related to the large-scale
behavior of Quantum ChromoDynamics (QCD). 
Remarkably, quarks and gluons appear to be confined in 
ordinary matter, due to the mechanism of color confinement which is not yet 
fully understood.
A detailed understanding of color confinement is one of the central 
goals of nonperturbative studies of QCD.
Lattice formulation of QCD allows us to investigate the color 
confinement phenomenon within a nonperturbative framework.
It is known since long that, in lattice numerical simulations,
tubelike structures emerge by analyzing the chromoelectric fields 
between static quarks~\cite{DiGiacomo:1990hc,Cea:1992sd,Matsubara:1993nq,Cea:1995zt,Bali:1994de,Cardaci:2010tb,Cea:2012qw,Cea:2014uja,Cardoso:2013lla}.
Such tubelike structures naturally lead to a linear potential between static 
color charges and, consequently, to a direct numerical evidence of color 
confinement. \\
To explore on the lattice the field configurations produced by 
a static quark-antiquark pair, the following connected correlation 
function~\cite{DiGiacomo:1990hc,Kuzmenko:2000bq} was used:
\begin{equation}
\label{rhoW}
\rho_W^{\rm conn} = \frac{\left\langle {\rm tr}
\left( W L U_P L^{\dagger} \right)  \right\rangle}
              { \left\langle {\rm tr} (W) \right\rangle }
 - \frac{1}{N} \,
\frac{\left\langle {\rm tr} (U_P) {\rm tr} (W)  \right\rangle}
              { \left\langle {\rm tr} (W) \right\rangle } \; ,
\end{equation}
where $U_P=U_{\mu\nu}(x)$ is the plaquette in the $(\mu,\nu)$ plane, connected
to the Wilson loop $W$ by a Schwinger line $L$, and $N$ is the number of colors
(see Fig.~\ref{fig:op_W}).
The quark-antiquark field strength tensor is given by~\cite{DiGiacomo:1990hc,Kuzmenko:2000bq}:
\begin{equation}
\label{fieldstrengthW}
F_{\mu\nu}(x) = \sqrt\frac{1}{g^2} \, \rho_W^{\rm conn}(x)   \; ,
\end{equation}
where $g$ is the gauge coupling.
In previous studies~\cite{Cea:1994ed,Cea:1995zt,Cardaci:2010tb,Cea:2012qw,Cea:2014uja,Cea:2015wjd} color flux tubes
made up of chromoelectric field directed along the line joining a static 
quark-antiquark pair have been investigated, in the cases of SU(2) and 
SU(3) pure gauge theories at zero temperature.
\\
In this paper  we study
the profile of color flux tubes in pure SU(3) gauge theory and in QCD with (2+1) flavors
and present some new preliminary results  with sources at distances up to 1.14 fm.
\\
The dual superconductor model~\cite{'tHooft:1976ep,Mandelstam:1974pi} of the QCD vacuum
is, at least, a very useful phenomenological frame to interpret the vacuum dynamics and the formation of
color flux tubes in the QCD vacuum.
In Refs.~\cite{Cea:2012qw,Cea:2014uja} it has been 
suggested that lattice data for chromoelectric flux tubes can be analyzed by 
exploiting the results presented in Ref.~\cite{Clem:1975aa}, where, from the 
assumption of a simple variational model for the magnitude of the normalized 
order parameter of an isolated vortex, an analytic expression is derived for 
magnetic field and supercurrent density, that solves the Ampere's law and the 
Ginzburg-Landau equations. As a consequence, the transverse distribution of  
the chromoelectric flux tube can be  described, according 
to~\cite{Cea:2012qw,Cea:2014uja,Cea:2015wjd}, by
\begin{equation}
\label{clem1}
E_l(x_t) = \frac{\phi}{2 \pi} \frac{1}{\lambda \xi_v} \frac{K_0(R/\lambda)}
{K_1(\xi_v/\lambda)} \;, \qquad  R=\sqrt{x_t^2+\xi_v^2}
\end{equation}
where $\xi_v$ is a variational core-radius parameter.
Equation~(\ref{clem1}) can be rewritten as
\begin{equation}
\label{clem2}
E_l(x_t) =  \frac{\phi}{2 \pi} \frac{\mu^2}{\alpha} \frac{K_0[(\mu^2 x_t^2 
+ \alpha^2)^{1/2}]}{K_1[\alpha]} \; , \quad \mu= \frac{1}{\lambda} \,, \quad \frac{1}{\alpha} =  \frac{\lambda}{\xi_v} \,.
\end{equation}
By fitting Eq.~(\ref{clem2}) to flux-tube data, one can get 
both the penetration length $\lambda$ and the ratio of the penetration length 
to the variational core-radius parameter, $\lambda/\xi_v$. Moreover,   
the Ginzburg-Landau $\kappa$ parameter can be obtained by
\begin{equation}
\label{landaukappa}
\kappa = \frac{\lambda}{\xi} =  \frac{\sqrt{2}}{\alpha} 
\left[ 1 - K_0^2(\alpha) / K_1^2(\alpha) \right]^{1/2} \,.
\end{equation}
Finally, the coherence length $\xi$ is determined by combining  
Eqs.~(\ref{clem2}) and~(\ref{landaukappa}).
\\
\begin{figure}[tb] 
\centering
\includegraphics[scale=0.4,clip]{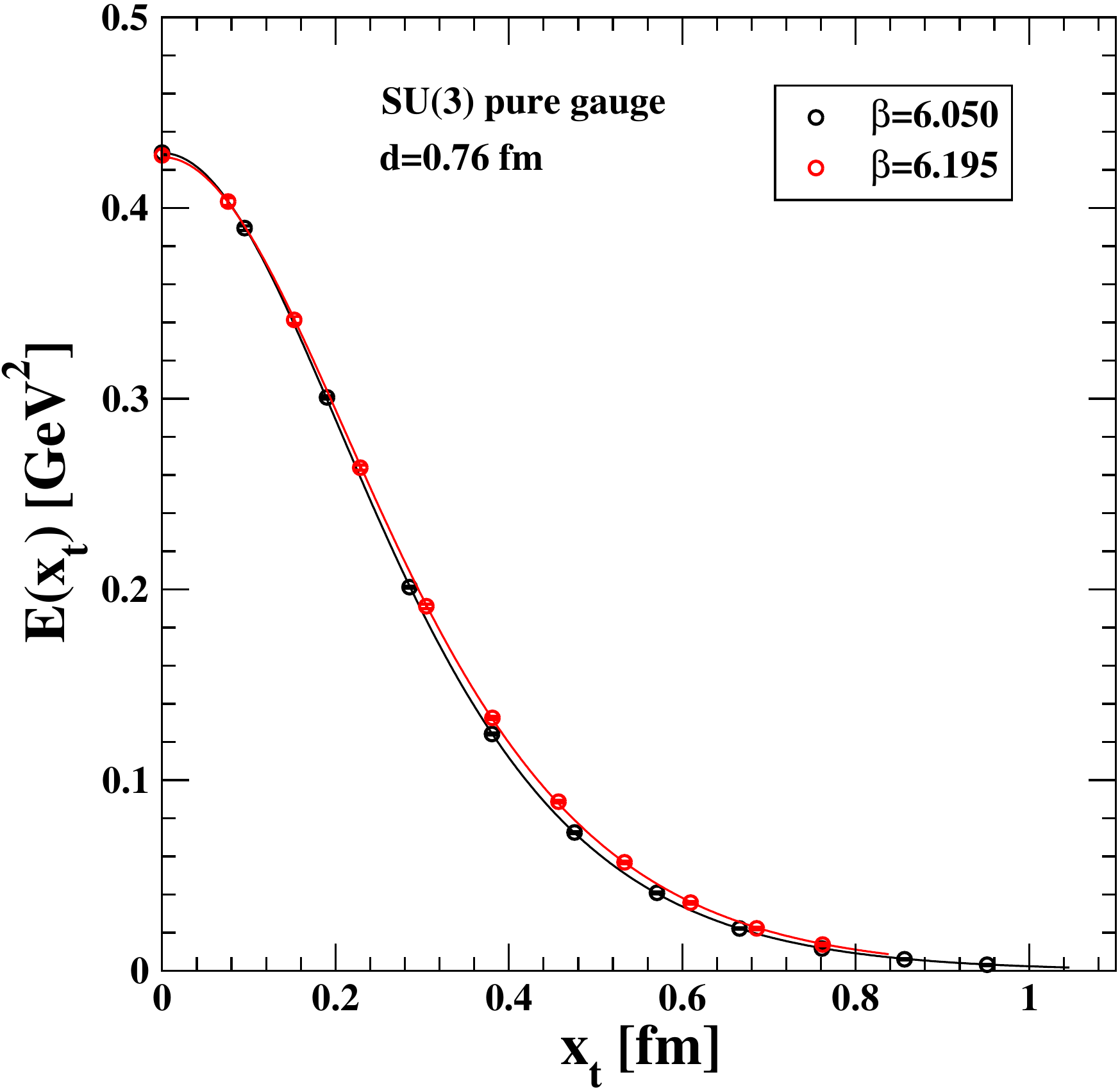}
\hspace{0.1cm}
\includegraphics[scale=0.4,clip]{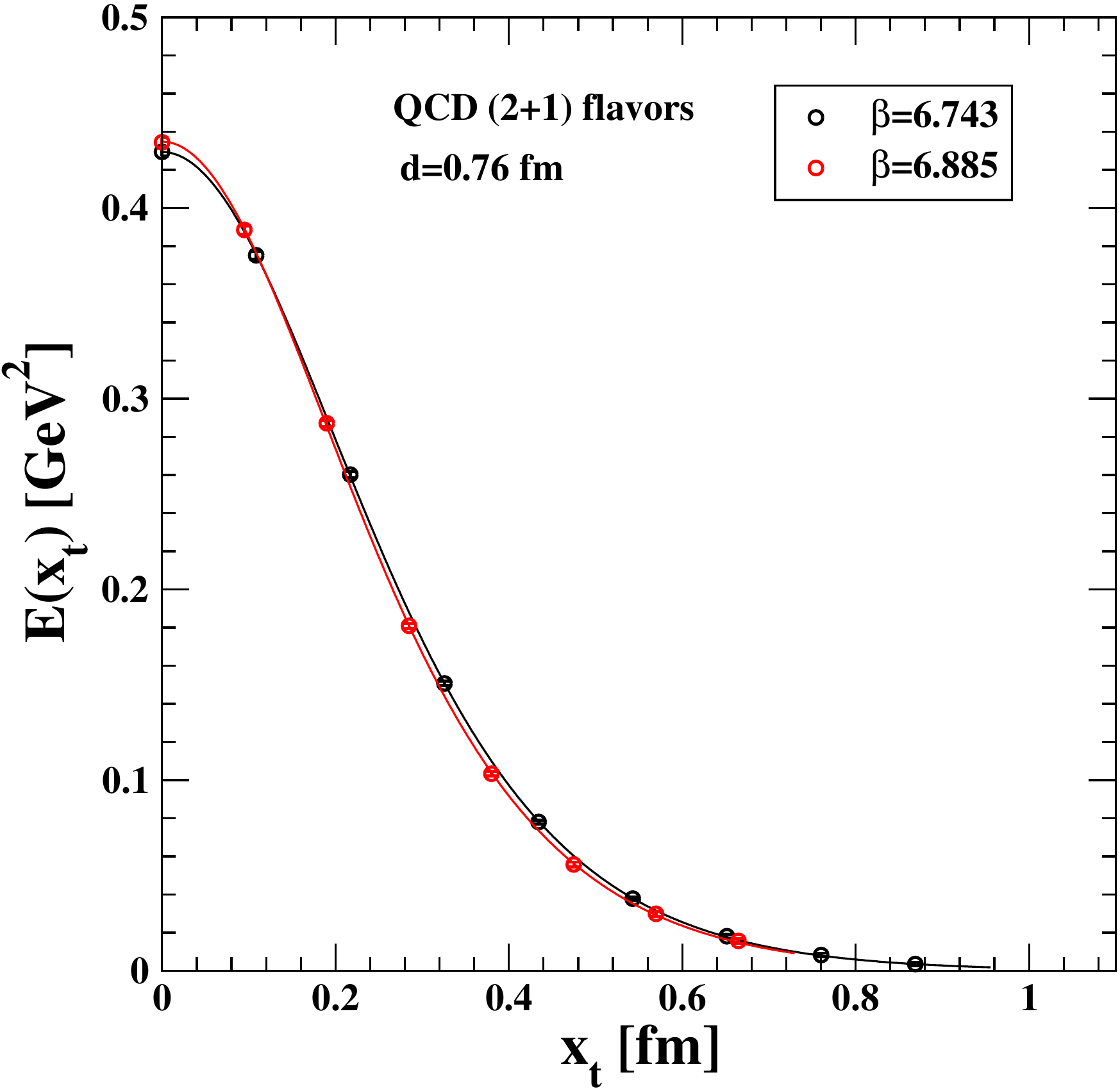}
\caption{(left) SU(3) pure gauge: the chromoelectric field $E_l(x_t)$ in physical units versus the transverse distance $x_t$ measured for two different values
of the gauge coupling for a distance $d=0.76 \,\,{\rm fm}$ between sources. Full lines are the fits using Eq.~(\ref{clem2}). 
(right) QCD (2+1) flavors: as for left figure.}
\label{fig:scalingSU3QCD}
\end{figure}

\section{Lattice setup and numerical results}
\label{setup}
Both for the cases of pure gauge SU(3) and QCD with (2+1) flavors we performed simulations
on a $32^4$ lattice. We have made use of the 
publicly available MILC code~\cite{MILC}, which has been suitably modified by 
us in order to introduce the relevant observables. 

\subsection{SU(3) pure gauge}
\label{SU3}

\begin{table}[thb]
\begin{center} 
\caption{SU(3) pure gauge. The results of the fit to the chromoelectric field Eq.~(\ref{clem2}) at several distances $d$ between the sources, together with
the square root width of the flux tube Eq.~(\ref{width}) and the square root of the energy in the flux tube per unit lenght normalized to the flux $\phi$,  Eq.~(\ref{energy}). }
\begin{tabular}{|c|c|c|c|c|c|c|c|c|c|c|}
\hline\hline
$\beta$ & $d$ [fm] & $\phi$ & $\lambda$ [fm]& $\kappa=\lambda/\xi$ & $\xi$ [fm] & $\sqrt{w^2}$ [fm]& $\sqrt{\varepsilon}/\phi$ [GeV]\\ \hline
6.050	&0.76	&5.143(39)	&0.164(5)  	&0.348(208)	&0.472(283)	&0.458(17)      &0.133(5)\\
6.195	&0.76	&5.485(56)	&0.173(7)  	&0.369(229)	&0.469(293)	&0.476(24)      &0.128(7)\\
6.050	&0.95	&5.932(114)	&0.169(16)  	&0.229(103)	&0.738(339)	&0.517(59)     &0.116(14)\\
6.050	&1.14	&6.254(617)	&0.166(95)  	&0.174(65)	&0.953(651)	&0.542(386)    &0.109(82)\\
\hline\hline 
\end{tabular} 
\end{center}
\end{table}

We measure on the lattice the chromoelectric field generated by a quark-antiquark pair at distances up to 1.14 fm.
The scale is fixed using the parameterization~\cite{Edwards:1998xf}:
\bea
\label{sqrt-sigma-SU3}
\left ( a \, \sqrt{\sigma} \right )(g) &=& f_{{\rm{SU(3)}}}(g^2) 
\left \{ 1+0.2731\,\hat{a}^2(g)  \right .\\
&-&0.01545\,\hat{a}^4(g) +0.01975\,\hat{a}^6(g) \left . \right \}/ 0.01364 \;  , \nonumber
\eea
\[
\hat{a}(g) = \frac{f_{{\rm{SU(3)}}}(g^2)}{f_{{\rm{SU(3)}}}(g^2(\beta=6))} 
\;, \;
\beta=\frac{6}{g^2} \,, \;\;\; 5.6 \leq \beta \leq 6.5\;,
\]
with
\beq
\label{fsun}
f_{{\rm{SU(3)}}}(g^2) = \left( {b_0 g^2}\right)^{- b_1/2b_0^2} 
\, \exp \left( - \frac{1}{2 b_0 g^2} \right) \;, \;\;\; b_0 \, = \, \frac{11}{(4\pi)^2} \; \; , \; \; b_1 \, = \, \frac{102}{(4\pi)^4} \,.
\eeq
In the following, we assumed for the string tension the standard value 
of $\sqrt{\sigma} = 420$ MeV. 
\\
We measured the connected correlator given in Eq.~(\ref{rhoW}) at 
the middle of the line connecting the static color sources, for various 
values of the distance between the sources and for integer transverse 
distances. 
In order to reduce the ultraviolet noise, we applied to the operator in 
Eq.~(\ref{rhoW}) one step of HYP smearing~\cite{Hasenfratz:2001hp} 
to temporal links, with smearing parameters $(\alpha_1,\alpha_2,\alpha_3) 
= (1.0, 0.5, 0.5)$, and $N_{\rm APE}$ steps of APE 
smearing~\cite{Falcioni1985624} to spatial links, with 
smearing parameter $\alpha_{\rm APE} = 0.40$. Here $\alpha_{\rm APE}$ is the ratio 
between the weight of one staple  and  the weight of the original link.

We fitted our data for the transverse 
shape of the longitudinal chromoelectric field to Eq.~(\ref{clem2}).  
Remarkably, we found that Eq.~(\ref{clem2}) is able to reproduce the transverse 
profile of the longitudinal chromoelectric field.

\begin{figure}[tb] 
\centering
\includegraphics[scale=0.4,clip]{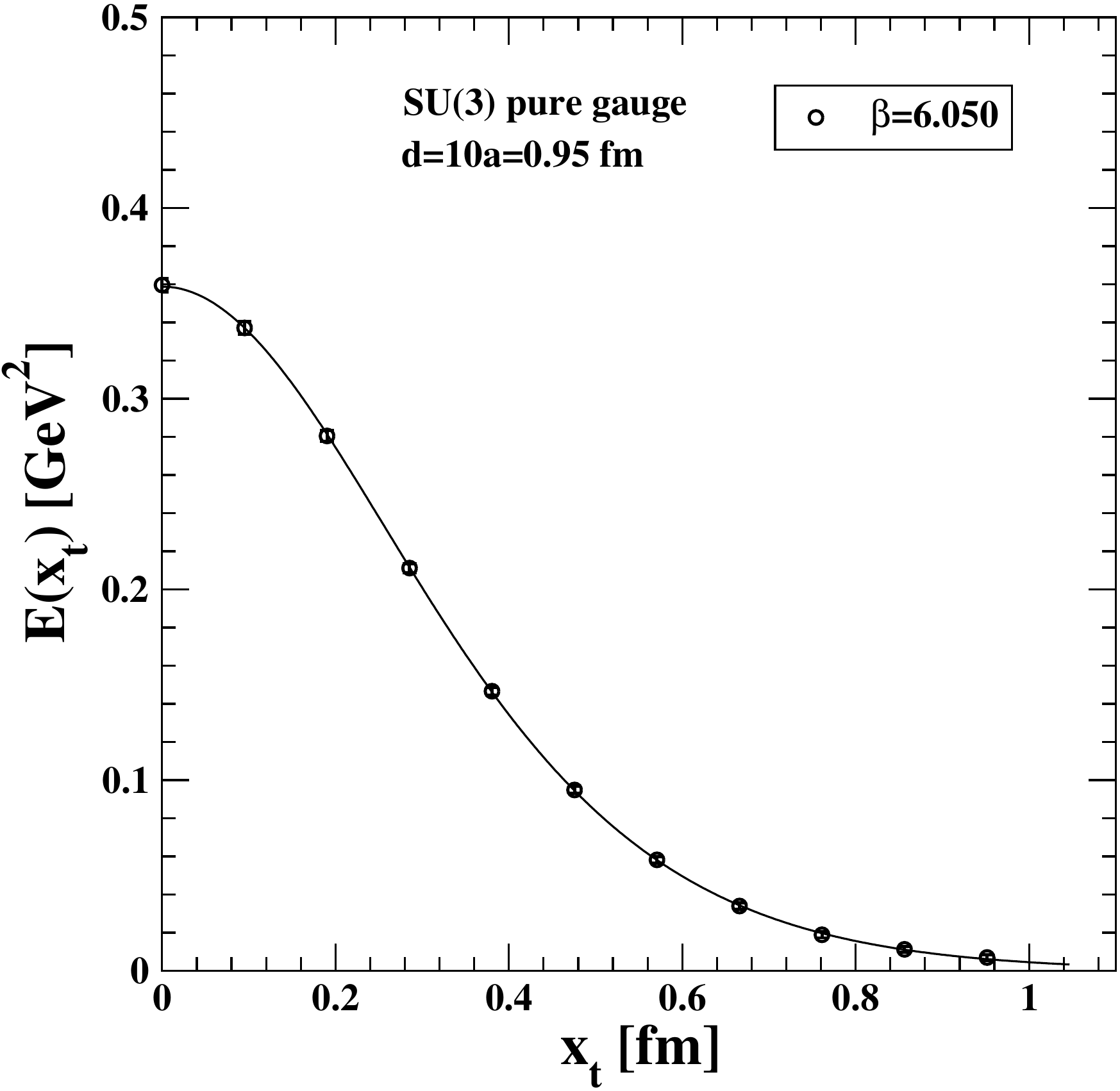}
\hspace{0.1cm}
\includegraphics[scale=0.4,clip]{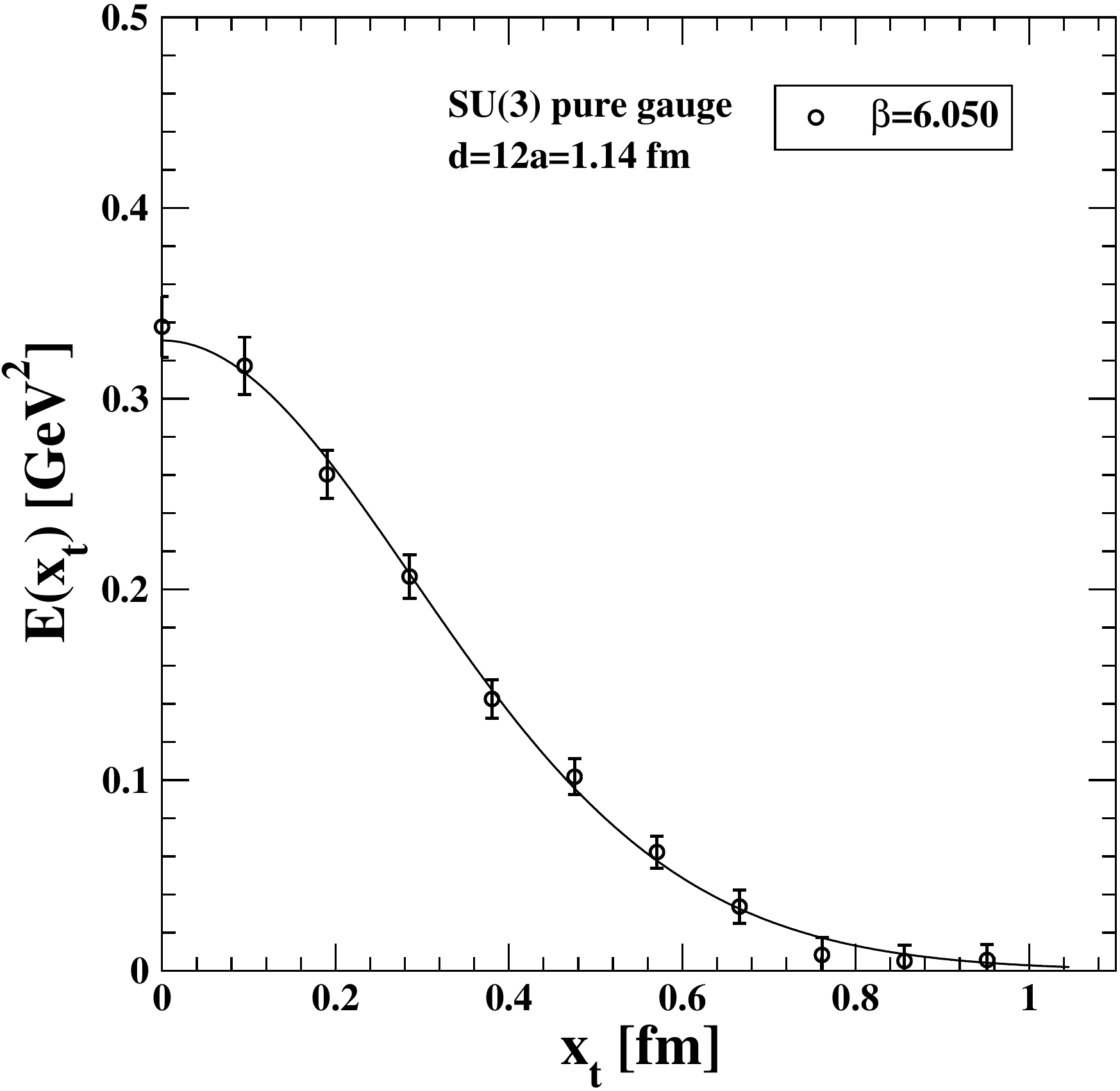}
\caption{SU(3) pure gauge: the chromoelectric field $E_l(x_t)$ in physical units versus the transverse distance $x_t$ measured for 
distance $d=0.95 \,\,{\rm fm}$ (left) and distance $d=1.14 \,\,{\rm fm}$ (right) between sources. Full lines are the fits using Eq.~(\ref{clem2}).}
\label{fig:fieldSU3}
\end{figure}

We checked that our lattice results are consistent with continuum scaling. To do this we
measured the longitudinal chromoelectric field generated by sources at distance 
$8a$ and $10a$ (($a$ is the lattice spacing) for two values of the gauge couplings 
$\beta=6.050$ and $\beta=6.195$. According to the scale given in Eq.~(\ref{sqrt-sigma-SU3}) this
amounts to a distance of $0.76 {\rm fm}$ in physical units. 
The result in Fig.~\ref{fig:scalingSU3QCD} seems to display an almost perfect scaling.
Having selected the gauge coupling region where continuum scaling holds, we measured the 
the longitudinal chromoelectric field at distances $10a$ and $12a$  at $\beta=6.050$ which corresponds
respectively to distances $0.95 {\rm fm}$ and $1.14 {\rm fm}$ in physical units.
In Fig.~\ref{fig:fieldSU3} we display the results.
Now, using Eq.~(\ref{clem2}) and the values of the parameters obtained by fitting Eq.~(\ref{clem2}) to
the numerical value for the longitudinal chromoelectric field, we are able to estimate the mean square root width of the 
flux tube:
\begin{equation}
\label{width}
\sqrt{w^2} = \sqrt{\frac{\int d^2x_t \, x_t^2 E_l(x_t)}{\int d^2x_t \, E_l(x_t)}} = \sqrt{\frac{2 \alpha}{\mu^2} \frac{K_2(\alpha)}{K_1(\alpha)}}
\end{equation}
and the square root of the energy in the flux tube per unit length, normalized to the flux $\phi$:
\begin{equation}
\label{energy}
\frac{\sqrt{\varepsilon}}{\phi} = \frac{1}{\phi} \sqrt{ \int d^2x_t \, \frac{E_l(x_t)^2}{2} } =  \sqrt{ \frac{\mu^2}{8 \pi} \, \left(1-\left(\frac{K_0(\alpha)}{K_1(\alpha)}\right)^2\right)}
\end{equation}
The results are given in Table~1.
We can argue that the penetration length $\lambda$ is almost stable within errors at varying the distance between the sources. While there is a hint of slow increasing for $\xi$ and 
$\sqrt{w^2}$.

\subsection{QCD (2+1) flavors}
\label{QCD}
\begin{figure}[tb] 
\centering
\includegraphics[scale=0.4,clip]{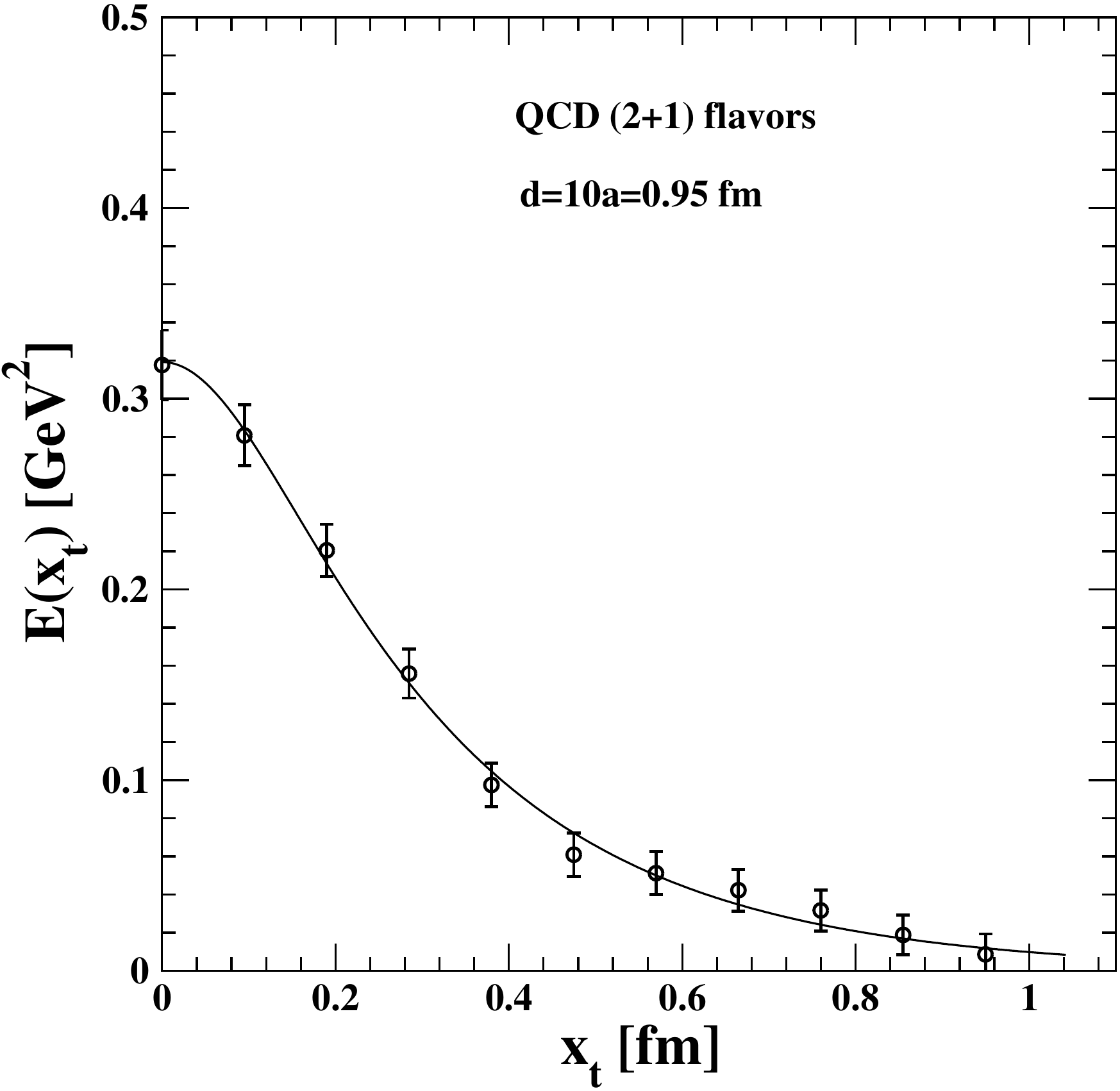} 
\hspace{0.1cm}
\includegraphics[scale=0.4,clip]{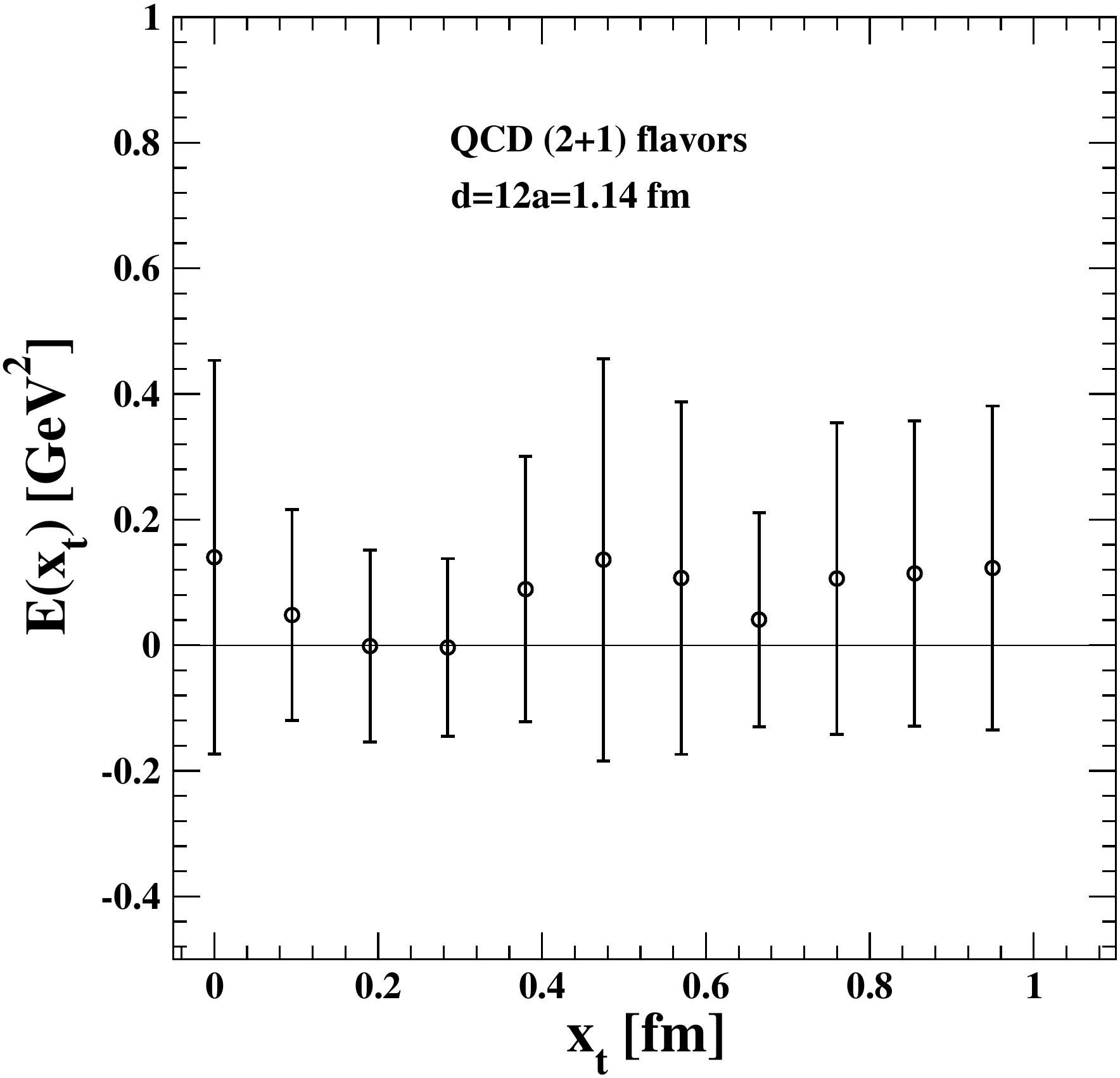}
\caption{QCD (2+1) flavors: the chromoelectric field $E_l(x_t)$ in physical units versus the transverse distance $x_t$ measured for 
distance $d=0.95 \,\,{\rm fm}$ (left) and distance $d=1.14 \,\,{\rm fm}$ (right) between sources. Full lines are the fits using Eq.~(\ref{clem2}).}
\label{fig:fieldQCD}
\end{figure}

In this section we present results obtained for QCD with (2+1) flavors.
Highly improved staggered quark action with tree level improved Symanzik gauge action (HISQ/tree) has been 
adopted~(see ref.()).
We  work on the line of constant physics determined by fixing the strange quark mass to its physical value $m_s$
at each value of the gauge coupling. The light-quark mass has been fixed at $m_l=m_s/20$. This
correspond to a pion mass $M_\pi=160 \,\, {\rm MeV}$.
The lattice spacing has been determined using results of Ref.~\cite{Bazavov:2011nk}.
As in the case of pure gauge SU(3) theory in order to measure the correlator Eq.~(\ref{rhoW}) we perform
one HYP smearing on temporal links and several APE smearings on spatial links.
To check the continuum scaling we considered two different values of the gauge couplings 
$\beta=6.743$ and $\beta=6.885$ and measured the chromoelectric field produced by sources
at distances $7a$ and $8a$ respectively. This amounts to have a distance of $0.76 \,\,{\rm fm}$ between sources.
The result displayed in Fig.~\ref{fig:scalingSU3QCD} indicates  a  almost perfect scaling (see also Table~2 for the parameters 
obtained in fitting the data to Eq.~(\ref{clem2}).
Then we measure the field produced for two other distances between the sources. Namely 
at gauge coupling $\beta=6.885$ we consider the distances $10a$ and $12a$ corresponding to
$0.95 \,\,{\rm fm}$ and $1.14 \,\,{\rm fm}$. The results are displayed in Fig.~\ref{fig:fieldQCD}.
At variance with respect to the pure SU(3) gauge (see Fig.~\ref{fig:fieldSU3}) the measurements for the chromoelectric field
at distance $d=1.14\,{\rm fm}$ versus the the transverse distance  seems to fluctuate around zero, although with large errors.
This circumstance could suggest that, in presence of dynamical fermions and for a  sufficiently large distance between sources,
the flux tube structure disappears.

\begin{table}[thb]
\begin{center} 
\caption{QCD (2+1) flavors. The results of the fit to the chromoelectric field Eq.~(\ref{clem2}) at several distances $d$ between the sources, together with
the square root width of the flux tube Eq.~(\ref{width}) and the square root of the energy in the flux tube per unit lenght normalized to the flux $\phi$,  Eq.~(\ref{energy}). }
\begin{tabular}{|c|c|c|c|c|c|c|c|c|c|}
\hline\hline
$\beta$ & $\Delta$ [fm] & $\phi$ & $\lambda$ [fm]& $\kappa=\lambda/\xi$ & $\xi$ [fm] & $\sqrt{w^2}$ [fm]& $\sqrt{\varepsilon}/\phi$ [GeV]\\ \hline
6.743	&0.76	&4.366(48)	&0.139(8)  	&0.264(131)	&0.526(263)	&0.411(29)     &0.146(11)\\
6.885        &0.76	&4.251(67)	&0.147(10)  	&0.342(203)	&0.429(256)	&0.411(33)     &0.148(13)\\
\hline\hline 
\end{tabular} 
\end{center}
\end{table}
\section*{Acknowledgements}

This work was based in part on the MILC Collaboration's public lattice gauge theory code ({\url{http://physics.utah.edu/~detar/milc.html}) and has been partially supported by INFN SUMA project.
Simulations have been performed using computing facilities at CINECA (INF16$\_$npqcd project under CINECA-INFN agreement).

\providecommand{\href}[2]{#2}\begingroup\raggedright\endgroup

\end{document}